\documentstyle[aps,prl,preprint]{revtex}

\begin{document}
\title{Anomalous diffusion as a signature of collapsing phase 
in two dimensional self-gravitating systems}
\author{Micka{\"e}l Antoni}
\address{Max-Planck-Institut f{\"u}r 
Physik Komplexer Systeme, Bayreuther Str. 40,
D-01187 Dresden (Germany)}
\author{Alessandro Torcini\thanks{INFM, Firenze (Italy)}}
\address{Dipartimento di Energetica "S. Stecco", Universit\`a di Firenze,
via S. Marta, 3 , I-50139 Firenze (Italy)}
\date{\today}

\draft
\maketitle
\begin{abstract}
A two dimensional self-gravitating Hamiltonian model made by $N$ fully-coupled 
classical particles exhibits a transition from
a collapsing phase (CP) at low energy to a homogeneous phase 
(HP) at high energy. From a dynamical point of view, the two 
phases are characterized by two distinct single-particle 
motions : namely, superdiffusive in the CP and ballistic 
in the HP. Anomalous diffusion is observed up to a time $\tau$
that increases linearly with $N$. Therefore, 
the finite particle number acts like a white noise 
source for the system, inhibiting anomalous
transport at longer times.
\end{abstract}

In the past years the thermodynamical properties of 
gravitatial models have been studied in detail from 
a theoretical \cite{thirr} and computational 
\cite{compagn,posch,martin} point of view. 
In particular, it has been shown that
at low energy the gravitational forces give rise to 
a collapsing phase (CP), identified by
the presence of a single cluster of particles floating 
in a diluted homogeneous background. At high energy a 
homogeneous phase (HP) is recovered:
the cluster disappears and the particles 
move almost freely. In the transition region the system 
is characterized (in the microcanonical ensemble) by a 
negative specific heat: the corresponding instability
(termed "gravo-thermal catastrophe") is of extreme
relevance for astrophysics (see \cite{lynden} for more
details). This apparent thermodynamical inconsistency has 
been solved by  Hertel and Thirring in Ref. \cite{thirr}, 
where they demonstrated the non equivalence of canonical and 
microcanonical ensemble in proximity of the transition region. 
These theoretical results have been successfully
confirmed by numerical investigations of self-gravitating
non-singular systems with short range interaction 
\cite{compagn,posch}.

More recently, in one dimensional lattices of fully 
and nearest-neighbour coupled symplectic maps 
with an attractive interaction it has been noticed that 
clustering phenomena are associated 
with anomalous diffusion (in particular, with subdiffusive
motion), at least for short times \cite{kaneko}.
Anomalous diffusion can be defined through the 
time dependence of the single particle mean square 
displacement (MSQD) $<r^2(t)>$, that typically reads as
\begin{equation}
< r^2(t) > \propto   t^\alpha  
\label{msqd}
\end{equation}
where the average $< \cdot >$ is performed over different
time origins and over all the particles of the system.
The transport is anomalous when $\alpha \ne 1$:
superdiffusive for $ 1 < \alpha < 2$ and subdiffusive
for $0 < \alpha < 1$ \cite{geisel2,grigo2}. Anomalous
transport has been revealed in dissipative
and Hamiltonian models \cite{geisel2} as well as
in experimental measurements \cite{solomon}. However,
the main part of literature focuses on systems with
few degrees of freedom (namely, one or two) and only few
studies have been devoted to extended models with $N \gg 1$ 
\cite{kaneko}.

In this Rapid Communication, the thermodynamical
and dynamical properties of a 2-D Hamiltonian 
system, constituted of $N$ particles interacting via a 
long range attractive potential, are analyzed.
In particular, we observe a transition from CP to 
HP associated to a dynamical transition 
from anomalous to ballistic transport.
Finite $N$ effects induce
a crossover from anomalous to normal diffusion at
long times. In the limit $N \to \infty$, the transport
mechanism remains anomalous for any time and reduces to
that of a single particle in an "egg-crate"
potential \cite{geisel}.

We consider a system of $N$ identical fully coupled particles 
with unitary mass evolving in a 2 dimensional periodic 
cell described by the Hamiltonian \cite{note0,ar}
\begin{equation}
H = K+V = \sum_{i=1}^N \frac{p_{x,i}^2+p_{y,i}^2}{2} 
+\frac {1}{2N} \sum_{i,j}^N
\Biggl[3-
\cos(x_i-x_j)
-\cos(y_i-y_j)-\cos(x_i-x_j)\cos(y_i-y_j)\Biggr] \quad ,
\label{ham}
\end{equation}
where $(x_i,p_{x,i})$ and $(y_i,p_{y,i})$ are the two pairs of 
conjugate variables with 
$(x_i,y_i) \in [-\pi,\pi[ \times [-\pi,\pi[$, 
$K$ and $V$ are the kinetic 
and potential energy, respectively. 
The potential part corresponds to the 
first three terms of the Fourier expansion of 
a 2-D attractive potential of the kind
$V(r) \propto \log|r|$. Such type of interaction
arises in self-gravitating 2-d gases
\cite{thirr,compagn,posch} as well as
in point vortices model for 2-d turbulence
\cite{frisch}. Due to the long range interaction 
among all the particles, this model can be described in terms 
of meanfield variables. In particular, the potential energy can be
rewritten as $V=\frac{1}{2}\sum_{i=1}^N V_i$, with
\begin{equation}
V_i = 3-M_x\cos(x_i-\phi_x)-M_y\cos(y_i-\phi_y) 
-\frac 1 2 
\Bigl[M^+_{xy}\cos(x_i+y_i-\phi^+_{xy})+ 
M^-_{xy}\cos(x_i-y_i-\phi^-_{xy}) \Bigr]
\label{ham1}
\end{equation}
where ${\bf M}_z = (<\cos(z)>_N,<\sin(z)>_N) = M_z {\rm exp}[i \phi_z]$
represents four two-dimensional meanfield vectors with
$z = x,y, x \pm y$ and $<..>_N$ denotes the average over $N$.
However, the single particle potentials $V_i$ are non-autonomous
since the meanfield quantities $M_z$ and $\phi_z$ are defined through 
the instantaneous values of the particles coordinates.
The motion of each particle is therefore determined self-consistently 
by an attractive and non-autonomous force field that is itself uniquely 
determined throught the motion of all the particles.
The self-gravitating nature of the model is due to this
effective force acting among the particles.
Since $V$ is invariant under the
transformations $x \leftrightarrow -x$, $y\leftrightarrow -y$ 
and $x \leftrightarrow y$, it turns out that in the "meanfield 
limit" (i.e. for $N \to \infty$ with $U=H/N$ constant) 
$M_x = M_y = M$ and $M^+_{xy}=M^-_{xy}=P$. Moreover, in this 
limit and assuming $\phi_z = 0$ the single potential $V_i$ turns out 
to be an egg-crate potential similar to that studied in \cite{geisel}. 
This periodic potential is characterized in each elementary cell by a minimum 
($V_m = 3 -2M -P$), 4 maxima ($V_M = 3 + 2M - P$)
and 4 saddle points ($V_s = 3+ P$).

For specific energy $U$ smaller than a critical value $U_c
\simeq 2$, we observe that the particles are mainly
in a clustered state. Above $U_c$ a HP is recovered. 
Following Refs. \cite{kaneko,ar}, the degree of 
clustering of the particles can be characterized 
through the time averages $<M_z>_t$ \cite{note1,future}.  
When at each time the particles have almost the same
position $<M_z>_t$ are ${\cal O}(1)$, while for a HP 
their values vanishes as $1/\sqrt{N}$ \cite{ar}. 
Fig.~1 shows that, for $U \to 0$, the average quantities
$M,P$ tend to one. This indicates that
the particles are almost all trapped in a potential 
well of depth $ \simeq (V_s - V_m)$ forming a compact cluster. 
For increasing energy $U$, the kinetic contribution becomes
more relevant and the average number of 
particles trapped in the potential well reduces.
As a consequence the value of $<M_z>_t$ decreases
together with $(V_s - V_m)$. For $U \ge U_c$, the
system is no more clustered and the particles
can move almost freely. Moreover, due to finite 
$N$ effects $<M_z>_t$ is not exactly zero, but 
${\cal O}(1/\sqrt{N})$.

In Fig.~2 the temperature $T = <K>_t/N$ is reported as 
a function of $U$. Above $U_c$, $T$ increases linearly with $U$ 
indicating that the system behaves like a free particle 
gas. In the CP, the tendency of the system to collapse
is balanced by the increase of the kinetic energy 
\cite{compagn}. This competition
leads initially (for $ 0 < U < 1.8$) to a steady
increase of $T$, followed (for $ 1.8 < U < U_c$) 
by a rapid decay of $T$. This yields a negative 
specific heat as illustrated in the inset 
of Fig.~2. These results are in full 
agreement with theoretical predictions based on the 
analysis of a simple classical cell model \cite{thirr} and 
with numerical findings \cite{compagn,posch} for short
ranged attractive potentials. 
The phenomenon of negative specific heat can be
explained within a microcanonical approach with
an heuristic argument \cite{thirr}: approaching
the transition, a small increase of $U$ leads
to a significative reduction of the number of collapsed
particles (as confirmed from the drop exhibited
by $M$ and $P$ for $U > 1.8$); as a consequence 
the value of $V$ grows and, due to
energy conservation, the system becomes cooler.

Our data confirm also another important prediction of 
Hertel and Thirring \cite{thirr}: the non-equivalence of 
canonical and microcanonical ensemble nearby the transition region.
In the inset of Fig.~2 are reported the microcanonical
findings, obtained via standard molecular dynamics (MD) simulations,
and the theoretical canonical results, derived in the
mean-field limit \cite{note,mac}. These two sets of
data coincide everywhere, except in the energy 
interval $1.6 < U < 2.0$.
The discrepancy is due to the impossibility of the canonical ensemble 
to exhibit a negative specific heat, a prohibition that does not hold for 
the microcanonical ensemble. Our theoretical estimation
of the Helmholtz free energy $F=F(T)$ reveals that usually 
$F$ has an unique minimum. For $T < 0.5$ the minimum $F_{C}$
corresponds to non zero values of $M$ and $P$ (i.e. to the CP),
while for $T > 0.55$ the minimum $F_{H}$ is associated
to $M=P=0$ (i.e. to the HP). In the region $0.5 < T < 0.55$,
both minima $F_C$ and $F_{H}$ coexist as local minima
of the free energy. However, for $T < T_c = 0.54$ 
the CP is observed because $F_C < F_{H}$ , while
for $T > T_c$ the HP prevails since $F_{H} < F_C$.
At $T=T_c$ the two minima are equivalent
and a jump in energy from $U(T_c^-) \approx 1.6$ to
$U(T_c^+) \approx 2.0$ is observed. This picture suggests
that this transition can be considered as a 
first order transition \cite{compagn}.

Let us now investigate if the observed thermodynamical
transition has any consequence on the dynamical behaviour
of the system. In order to characterize the single particle
dynamics, we consider the MSQD $<r^2(t)>$. 
As shown in Fig.~3, in the CP the diffusion is anomalous 
for times shorter than a 
crossover time $\tau$, while for longer 
times the Einstein law is recovered $<r^2(t)> \propto 4 D t$ 
(where $D$ is the diffusion coefficient).
A similar behaviour for the MSQD has been already 
observed for a system of $N$ coupled symplectic maps 
in \cite{kaneko}, but with $\alpha < 1$. However,
in the present case $\tau$ increases linearly
with $N$ \cite{future} indicating that
in the meanfield limit the asymptotic dynamical regime 
will be superdiffusive \cite{tau}.

The observed dynamical behaviour can be explained
noticing that in the mean-field limit each particle $i$
will see essentially the same constant 2-D egg-crate
potential $V_i$. Moreover, it has been shown 
in Ref. \cite{geisel} that a single particle moving 
in a egg-crate potential with an energy between 
$V_s$ and $V_M$ exhibits superdiffusion. 
This is due to the fact that the particle moves
for long times almost freely along the channels of 
the potential and episodically is trapped for a while
in the potential well. In phase space, the superdiffusive 
phenomenon can be explained by a trapping mechanism in 
a hierarchy of cantori around a cylindrical KAM-surface 
\cite{geisel2}. Therefore, in our model (\ref{ham})
for $N \to \infty$ anomalous transport is due 
only to the fraction of particles that can move
along the channels.

For finite $N$, the potential $V_i$ seen by the particle $i$
will fluctuate in time. Hence, particles having an energy
close to $V_S$ have the possibility to be trapped in the
potential well as well as to escape from it. As a consequence,
for sufficiently long time scales each particle can
experience free and localized motions.
The fluctuations of the potential $V_i$ reflect themselves 
on the structure of the phase space, introducing a white noise that 
destroys the self-similar structure of the island 
chains and of the cantori below a certain cut-off size. 
Being the self-similarity no more complete, one expect
that on long time scales normal diffusion will
be recovered \cite{kaneko2}.

As pointed out in Refs. \cite{grigo1,grigo}, if white noise is
added to a dynamical system exhibiting superdiffusive 
behaviour, $D$ (measured in the limit $t \to \infty$)
is inversely proportional to the noise amplitude.
Therefore, we expect that in our model the
value of $D$ will increase with $N$. That is indeed 
the case, and we observe a power-law dependence 
of the type $D \propto N^\gamma$. For example, considering 
systems with $ 100 \le N \le 10,000$ we have found
for $U=1.48$ and $U = 2.00$ a $\gamma$-value equal to
$0.7 \pm 0.1$ and $1.0 \pm 0.1$, respectively. 
The $N$-dependence of the diffusion coefficient
can be explained noticing that 
$D \propto \tau^{\alpha-1}$ \cite{vacf}.
This result coincide with that found theoretically in
Ref.~\cite{grigo1} and confirmed numerically 
by considering as dynamical models two very simple noisy 
maps.
For subdiffusive motion $D$ is inversely proportional
to $\tau$ (as found in \cite{kaneko2}), while 
for superdiffusive motion
($\alpha > 1$) a direct proportionality is expected \cite{grigo1}.
As already reported $\tau \propto N$, therefore we will have that
$\gamma = \alpha -1$. 
Assuming for $\alpha$ the corresponding asymptotic values \cite{nfin}, 
we can estimate as theoretical values $\gamma \simeq 0.64$ 
and $\simeq 0.9$ for $U=1.48$ and 2.00, respectively. 
In view of all the present limitations, 
these values can be considered consistent with the numerical findings.

As a final point, we examine the dependence of the asymptotic
$\alpha$-values from the energy $U$ of the system. 
A transition from anomalous
diffusion to ballistic motion ($\alpha=2$) at $U \simeq U_c$ 
is evident from Fig.~1, where the $\alpha$-values, obtained for
$N=4,000$, are reported. In particular, for $0.4 \le U \le 2.0$, 
we observe an increase of $\alpha$ from $1.3\pm 0.1$ to $1.9 \pm 0.1$. 
This phenomenon is a consequence of the flattening of the single particle
potential (i.e. of the reduction of $V_M - V_m$) observed for growing
$U$. The decrease of the average number of particles trapped in the cluster,
and the consequent increase of those moving freely, naturally drives
the diffusion mechanism toward a ballistic behaviour.
Moreover, for $U > U_c$, the potential $V_i$ 
fluctuates with typical amplitude ${\cal O}(1/\sqrt{N})$
around a constant value and a ballistic motion is expected
for all the particles. For small energies ($U < 0.3$) the 
MSQD seems to saturate to a constant value, indicating 
that all the particles are always clustered.
However, we believe that in these cases our observation
time was not sufficient to detect particles escaping from the
potential well.

In conclusion, we have shown for the first time 
a thermodynamical transition associated to a 
dynamical transition from anomalous to ballistic
transport. Moreover, the transport in our 
$N$-body system can be interpreted in terms of a noisy 
single particle motion in a 2-D Hamiltonian egg-crate potential.
The asymptotic dynamics of the model is strongly influenced
by the order in which the two limits
$N \to \infty$ and $t \to \infty$ are taken.
Indeed, if the limit $N \to \infty$ is performed before
the limit $t \to \infty$ the diffusion will be
always anomalous. Otherwise normal diffusion is
recovered for sufficiently long times.

As noticed in \cite{thirr}, the dimensionality of the 
system should not affect the main characteristics 
of the observed thermodynamical transition. We expect 
that the same should be true also for the corresponding 
dynamical behaviours.
Moreover, anomalous diffusion should be observable
for atomic clusters \cite{lebastie}, turbulent 
vortices \cite{frisch} (for which indeed has been already
observed \cite{solomon}) and gravitational systems
\cite{lynden}. All the cited systems share as a
common aspect to exhibit a clustered phase.

As a final remark, we claim that the inclusion of other 
terms of the Fourier expansion in the expression of the 
gravitational potential, will not affect the main results
here presented. However, as far as transport properties are 
concerned, we believe the range of the force to play an 
important role. In particular, for short-ranged interactions
we expect a more chaotic behaviour of the particles. 
This may prevent the anomalous diffusion regime to persist
in the thermodynamical limit \cite{kaneko2}.
Future work will be devoted to the study of these
fundamental points \cite{future}.

We would like to thank L. Casetti, R. Livi, P. Grassberger, 
P. Grigolini, H. Kantz, J. Klafter, R. MacKay, S. Ruffo, A. Seyfried 
and G. Zumofen for useful suggestions and the University of
Wuppertal for the kind hospitality.
This work was also partially supported through the european 
contract N. ERBCHRX-CT94-0460.

\begin{figure}
\vskip 0.5truecm
\caption{
Time averages of $M$ and $P$ as a function of $U$. The 
solid curves refer to the theoretical estimation 
(i.e. to canonical results) and the symbols to the MD findings 
(i.e. to microcanonical results). 
The exponents $\alpha$, defined in eq. 
(\protect{\ref{msqd}}), are also reported 
(triangles). The MD data have been obtained with $N=4,000$ (apart 
few points with $N=10,000$) and averaged over a 
total integration time ranging from $t = 1,2 \times 10^6$ to 
$t=2,4 \times 10^6$, with a time step $dt = 0.3$. 
The $\alpha$-values have been estimated in the time
interval $150 < t < 10,000$ for any reported $U$.
}
\label{magnetiz}
\end{figure}

\begin{figure}
\vskip 0.5truecm
\caption{Temperature $T$ as a function of the specific energy 
$U$. The solid line corresponds to the analytical estimation
(canonical ensemble) and the triangles to the simulations results 
(microcanonical ensemble).  In the inset, an enlargment of the 
transition region is reported: the solid (resp. dashed) curves 
refer to the principal (resp. relative) minimum of $F(T)$.
The parameters for the MD simulations are the same as 
in Fig.~1.
}
\label{TvsU}
\end{figure}

\begin{figure}
\vskip 0.5truecm
\caption{
Mean square displacement $<r^2(t)>$ as a function of time in 
a log-log plot. The cross-over time $\tau$ from anomalous to normal diffusion
is also reported. The data refer to $U=1.1$ and $N=4,000$, the total 
integration time is $t = 1.2 \times 10^6$. 
In the inset, the logarithm of the 
cross-over time $\tau$ is reported as a function of $\ln(N)$ for $U=1.48$. 
The reported $\tau$-values (circles) have been estimated adopting a 
threshold $\beta=1.1$ (for more details see [17]). 
The solid line represents
a best linear fit to the data and its slope is $0.95 \pm 0.08$.
}
\label{MvsT}
\end{figure}


\begin{references}

\bibitem{thirr} P. Hertel and W. Thirring, Ann. of Physics, {\bf 63},
520 (1970).

\bibitem{compagn} A. Compagner, C. Bruin and A. Roesle,
Phys. Rev. A, {\bf 39}, 5989 (1989).

\bibitem{posch} H.A. Posch, H. Narnhofer and W. Thirring, Phys. Rev.
A {\bf 42}, 1880 (1990).

\bibitem{martin} P.A. Martin and J. Piasecki, J. Stat. Phys. {\bf 84},
837 (1996).

\bibitem{lynden} D. Lynden-Bell and R. Woo, Mon. Notic. Roy. Astron.
Soc. {\bf 138}, 495 (1968).

\bibitem{kaneko} K. Kaneko and T. Konishi, Physica D {\bf 71},
146 (1994).

\bibitem{geisel2} T. Geisel, in {\it Levy flights and related
topics in physics} , eds. M.F. Schlesinger et al. 
(Springer, Berlin, 1995) p. 153.

\bibitem{grigo2} P. Grigolini, in {\it Chaos: the interplay between
stochastic and deterministic behaviour} , eds. P. Garbaczewski et al. 
(Lecture Notes in Physics, Springer , Berlin, 1995) {\bf 457} p. 101.

\bibitem{solomon} T.H. Solomon, E.R. Weeks and H.L. Swinney, Phys.
Rev. Lett. {\bf 71}, 3975 (1993).

\bibitem{geisel} T. Geisel, A. Zacherl and G. Radons, Phys. Rev.
Lett. {\bf 58}, 1100 (1987) and Z. Phys. B,
Cond. Matter, {\bf 71}, 117 (1988); 
D.K. Chaikovsky and G.M. Zalavsky, Chaos, {\bf 1}, 463 (1991);
J. Klafter and G. Zumofen,
Phys. Rev. E, {\bf 49}, 4873 (1994).

\bibitem{note0} The present model can be considered as an
extention to 2-D of the model recently introduced
in \protect{\cite{ar}}.

\bibitem{ar} M. Antoni and S. Ruffo, Phys. Rev. E {\bf 52}, 2361 (1995).

\bibitem{frisch} U. Frisch, "Turbulence" (Cambridge University Press,1995);
T.S. Lundgren and Y.B. Pointin, J. Stat. Phys.  {\bf 17} (1977) 323.

\bibitem{note1} For finite values of $N$, the equalities 
$<M_x>_t = <M_y>_t =M$ and $<M^+_{+xy}>_t=<M^-_{xy}>_t=P$ 
are not expected to hold a priori, even in the limit $t \to \infty$. 
However, numerical tests indicate that the values of $<M_z>_t$ 
(averaged for a sufficiently long time) do not depend remarkably
from $N$. Apart exactly at the transition point $U = U_c$, where 
metastable states have been observed \protect{\cite{future}}. 
This justifies the identification of the time averages with
the mean field asymptotic values. 

\bibitem{future} A. Torcini and M. Antoni, in preparation.

\bibitem{note} The MD simulations have been performed
within the microcanonical ensemble adopting a 4th-order 
symplectic integrator \protect{\cite{mac}} with an 
integration time step $dt = 0.3$. The energy is conserved with a
relative precision of $10^{-8}$ for any considered $N$.
The canonical results have been obtained theoretically in the limit 
$N \to \infty$ from a straightforward estimation of the partition 
function obtained applying the Hubbard-Stratonovich trick,
analogously to what made in \protect{\cite{ar}}
(more details will be reported in \protect{\cite{future}}).

\bibitem{mac} R. I. McLachlan and P. Atela, Nonlinearity {\bf 5}, 541 (1992).

\bibitem{tau} In order to give an unambiguous definition of $\tau$,
we have considered the local slope of $\ln[<r^2(t)>]$ as a function
of $\ln(t)$. The crossover time is defined as the time for which
this local slope becomes smaller than a given threshold $\beta$.
We have verified the linear dependence of $\tau$ versus $N$ considering
two $\beta$-values (namely, $\beta = 1.1$ and $1.2$) and two energy
values (namely, $U=1.48$ and $2.00$). The dependence of $\tau$ on
$U$ is not monotonous, however we observe that for $N = 4,000$
$20,000 < \tau < 100,000$ choosing $\beta = 1.1$.

\bibitem{kaneko2} K. Kaneko and T. Konishi, Phys. Rev. A {\bf 40}, R6130
(1989).

\bibitem{grigo1} E. Floriani, R. Mannella, and 
P. Grigolini, Phys. Rev. E {\bf 52}, 5910 (1995).

\bibitem{grigo} R. Bettin, R. Mannella, B.J. West and 
P. Grigolini, Phys. Rev. E {\bf 51}, 212 (1995).

\bibitem{vacf} We have verified that the velocity 
autocorrelation function (VACF) shows a long-time tail decaying 
as $\propto t^{\beta}$, where $\beta \simeq \alpha -2$: 
a results fully consistent with the observed time dependence 
of the MSQD \protect{\cite{grigo2}}. However, for finite $N$ the 
VACF after a time $t_V$ exponentially vanishes. It is reasonable
to expect that $t_V \propto \tau$, by making this assumption 
the reported relation between
$D$ and $\tau$ is easily derived \protect{\cite{future}}.

\bibitem{nfin} 
For the present model (\protect{\ref{ham}}) 
a dependence of $\alpha$ on $N$ has been
noticed. However, a saturation to an
asymptotic value is clearly observed
for $N > 4,000$ \protect{\cite{future}}.

\bibitem{lebastie} "Clusters of atoms and molecules I",
ed. H. Haberland (Springer, Berlin, 1995);
P. Lebastie and R.L. Whetten, \prl
{\bf 65} (1990) 1567.

\end{references}
\end{document}